\documentclass[aps,prl,twocolumn,superscriptaddress,showpacs,amsfonts,amsmath,amssymb]{revtex4-1}
\usepackage{graphicx}
\usepackage{color}

\begin{document}

\title{Localized in-gap state in a single electron doped Mott insulator}

\author{Weng-Hang Leong}
\affiliation{Department of Physics and National Laboratory of Solid State Microstructure, Nanjing University, Nanjing 210093, China}

\author{Shun-Li Yu}
\affiliation{Department of Physics and National Laboratory of Solid State Microstructure, Nanjing University, Nanjing 210093, China}

\author{T. Xiang}
\affiliation{Institute of Physics, Chinese Academy of Sciences, P.O. Box 603, Beijing 100190, China}
\affiliation{Collaborative Innovation Center of Quantum Matter, Beijing, China}

\author{Jian-Xin Li}
\affiliation{Department of Physics and National Laboratory of Solid State Microstructure, Nanjing University, Nanjing 210093, China}

\date{\today}

\begin{abstract}
Motivated by the recent atomic-scale scanning tunneling microscope (STM) observation for a spatially localized in-gap state in an electron doped Mott insulator, we evaluate the local electronic state of the Hubbard model on the square lattice using the cluster perturbation theory. An in-gap state is found to exist below the upper Hubbard band around the dopant lattice site, which is consistent with the STM measurements. The emergence of this local in-gap state is accompanied with a rapid reduction of the double occupancy of electrons. A similar in-gap state is also found to exist on the triangular lattice. These results suggest that the in-gap state is an inherent feature of Mott insulators independent of the lattice structure.
\end{abstract}


\maketitle

The mechanism underlying high-$T_c$ superconductivity in cuprates remains one of the most challenging and fundamental problems in condensed matter physics~\cite{PA.Lee}. All cuprate superconductors have a layered structure made up of one or more CuO$_2$ planes. Their parent compounds have one unpaired electron per Cu unit cell, which constitutes a Mott insulating ground state with an antiferromagnetic (AF) long range order~\cite{PW.Anderson}. By doping holes or electrons to CuO$_2$ planes, the AF order is suppressed and a superconducting phase emerges above a critical doping concentration. The evolution from the AF Mott insulating phase to the superconducting phase induced by doping is highly non-trivial~\cite{PA.Lee}.
A thorough investigation on the charge and spin dynamics of doped Mott insulators is believed to be the key for the understanding of extraordinary phenomena observed in high-$T_c$ superconductors, such as the pseudogap effect.

The dynamics of a single hole in an antiferromagnetic Mott insulator has been extensively studied using the self-consistent Born approximation~\cite{CM.Varma, CL.Kane, F.Marsiglio, P.Hrsch, JM.Wheatley, J.Bala}, finite-size exact diagonalization~\cite{E.Gagliano,ZP.Liu,RJ.Gooding} and quantum Monte Carlo~\cite{EDagotto,TK.Lee} methods, based on the $t-J$-type model. It was predicted that the single-particle spectrum consists of a sharp coherent peak, corresponding to a quasiparticle excitation, and an incoherent background. But this sharp coherent peak was not observed in the spectrum of electrons measured by angle-resolved photoemission spectroscopy on Sr$_2$CuO$_2$Cl$_2$\cite{RJ.Birgeneau} and Ca$_2$CuO$_2$Cl$_2$~\cite{ZX.Shen,KM.Shen}. To reconcile the difference between theory and experiments, two kinds of scenarios were proposed to explain why the sharp quasipartilcle peak is absent. One is to attribute this absence of sharp peak as an extrinsic effect induced by electron-phonon coupling~\cite{Mishchenko}. The other regards this as an intrinsic effect resulting from a self-localization of doped holes in an AF background which smears out the coherent peak. From the self-consistent mean-field approximation (SCMFA), indeed it was found that the charge excitations are self-localized in a staggered AF ordered state~\cite{WP.Su, AR.Bishop, G.Seibold}. This kind of charge self-localization was also predicted to exist in a single-hole Hubbard model by considering the non-perturbative phase string effect~\cite{ZY.Weng} or in the underdoped Mott insulator when the chemical potential lies within the pseudogap~\cite{P.Phillips}.

Recently, local electronic structures in both doped and undoped Mott insulators were measured by STM~\cite{O. Fischer,YY.Wang}. In particular, the upper and lower Hubbard bands, which are particle-hole symmetric and spatially uniform, were observed in the recent STM study for the Mott insulator  Ca$_2$CuO$_2$Cl$_2$~\cite{YY.Wang}. Moreover, a broad local in-gap state whose energy is below the upper Hubbard band was observed around a Cl defect on the surface of Ca$_2$CuO$_2$Cl$_2$, which is effectively a single electron doped Mott insulator~\cite{YY.Wang}.

To understand the STM experimental results, we calculate the local spectral function of electron for the Hubbard model without and with one electron doping on both square and triangular lattices using the cluster perturbation theory (CPT)~\cite{M.Pioro, D.Plouffe}. We find that, apart from the Mott gap, a localized in-gap state emerges below the upper Hubbard band. The total spectral weight of this in-gap state shows a rapid increase with the Hubbard interaction in a regime where the double occupation number of electrons shows a sharp reduction. Our result reveals the physical origin of this in-gap state and gives a natural account for the experimental observation~\cite{YY.Wang}. We also compare the result with that obtained using SCMFA and find that the two methods agree qualitatively with each other, but SCMFA overestimates the spectral weight for the local in-gap state.

The Hubbard model is defined by the Hamiltonian
\begin{equation}\label{eqhub}
H = -t\sum_{\langle i,j\rangle \sigma}c^{\dag}_{i\sigma}
    c_{j\sigma} + U\sum_{i}n_{i\uparrow}n_{i\downarrow} ,
\end{equation}
where $c^{\dag}_{i\sigma}$ is a creation operator of electron with spin $\sigma$ on site $i$ and $n_{i\sigma}=c^{\dag}_{i\sigma}c_{i\sigma}$. $\langle i,j\rangle$ represents a pair of the nearest neighbor sites.  The hopping constant $t$ is set to 1 in the discussion below.

\begin{figure}
\begin{center}
\includegraphics[width=0.44\textwidth]{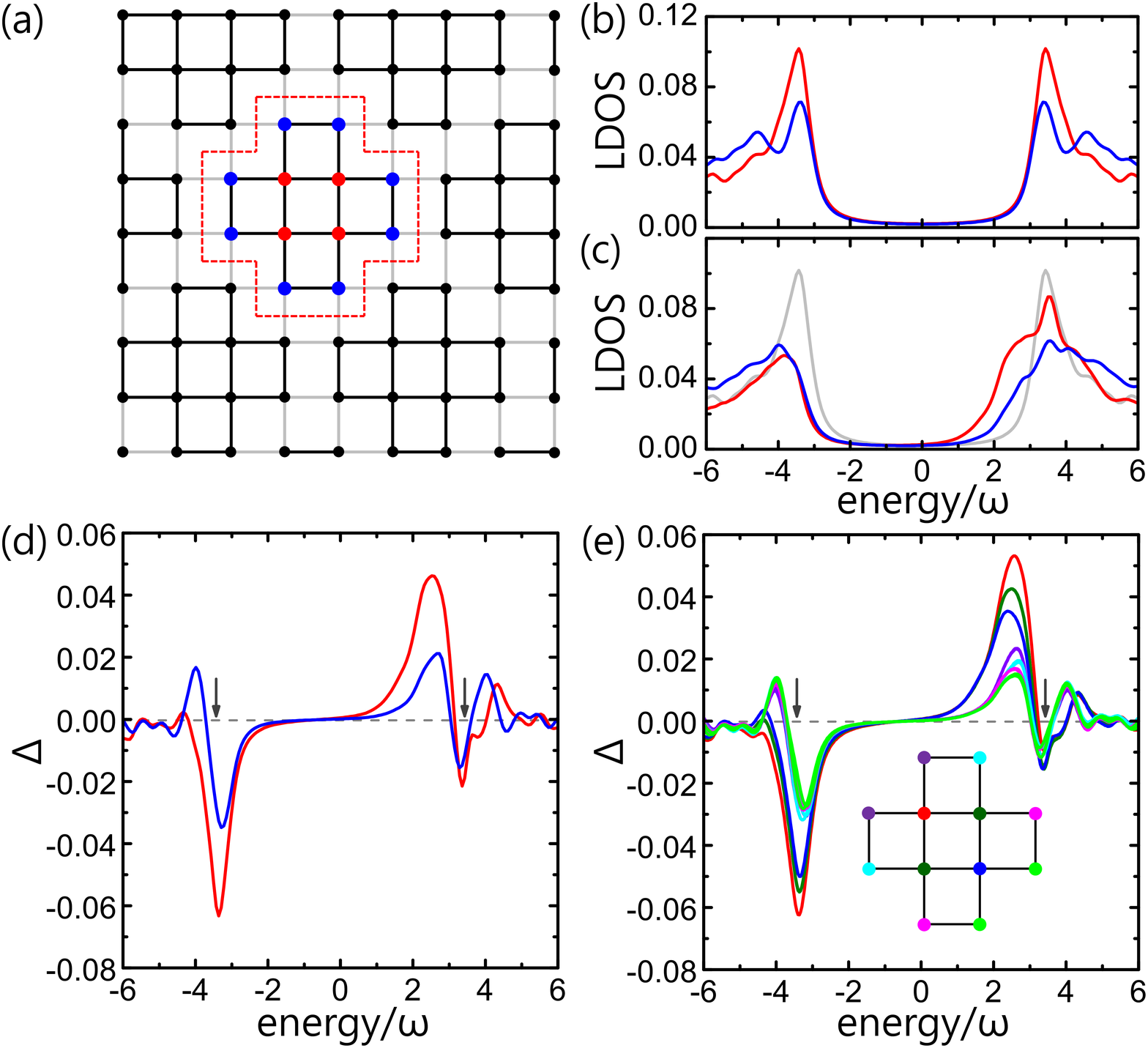}
\caption{(color online)
(a) A 12-site cluster, as enclosed by the red dashed lines, on the square lattice.
The black and grey lines denote the intra- and inter-cluster hoppings, respectively.
(b) and (c) show the LDOS within the clusters without electron doping (i.e. at half-filling) and with one electron doping, respectively. The color lines indicate the results on the sites with the same colors in Fig.(a).
The LDOS on the red sites in the undoped system is also shown with grey line in (c) for comparison.
(d) The difference $\Delta$ between the LDOS with and without electron doping on the corresponding color sites.
(e) $\Delta$ on sites indicated by color dots in the inset with an on-site attractive potential $v=-0.08$ on the red site.
The grey arrows in (d) and (e) indicate the peak positions of the LDOS for the lower and upper Hubbard bands on the red site at half filling.
}\label{fig1}
\end{center}
\end{figure}

The CPT is based on the exact diagonalization of finite clusters and the inter-cluster coupling is taken as perturbation~\cite{M.Pioro, D.Plouffe, S.Pairault,Zacher,Tremblay,Kang,Shun1,Shun2}. It can be used to treat a system with much larger lattice size than the exact diagonalization method.
In this method, the lattice is divided into a superlattice of decoupled clusters and each cluster contains $N$ sites (Fig.~\ref{fig1}(a)). In this superlattice, a lattice site is denoted by two indices $(\alpha, m)$, where $\alpha$ is the index of the cluster and $m$ is the lattice coordinate.
The Hamiltonian for the $\alpha$'th cluster $H_\alpha$ is first diagonalized using the Lanczos algorithm to obtain its ground-state energy $E^{\alpha}_{0}$ and the corresponding wavefunction  $|\Omega^{\alpha}\rangle$.
From this, the single-particle Green's function is calculated,
\begin{eqnarray}
G''^{\alpha}_{m\sigma,m'\sigma'}(z)& = & \langle\Omega^{\alpha}| c^{\alpha}_{m\sigma}\frac{1}{z-H_{\alpha}+E^{\alpha}_{0}} c^{\alpha\dag}_{m'\sigma'}|\Omega^{\alpha}\rangle \nonumber \\
& + & \langle\Omega^{\alpha}|c^{\alpha\dag}_{m\sigma} \frac{1}{z-H_{\alpha}-E^{\alpha}_{0}}c^{\alpha}_{m'\sigma'}| \Omega^{\alpha}\rangle,
\end{eqnarray}
where $z$ is a complex frequency.

The hopping between different clusters is taken as perturbation. We evaluate the local spectrum of electron in real space. In this case, the Green's function for the whole system, which is a $2N_{s}\times 2N_{s}$ matrix with $N_s$ the number of the whole lattice sites, can be expressed as
\begin{equation}
\textbf{G}_{\mathrm{cpt}}^{-1}(z)=\textbf{G}'^{-1}(z)-\textbf{V},
\end{equation}
where the matrix elements of $\textbf{G}'(z)$ and the intercluster hopping term $\textbf{V}$ are defined by
\begin{eqnarray}
&& G'_{\alpha m\sigma, \beta m'\sigma'} (z) = \delta_{\alpha\beta} G''^{\alpha}_{m\sigma,m'\sigma'}(z) , \\
&& V_{\alpha m\sigma,\beta m'\sigma'} = - t \delta_{\alpha \not= \beta} \delta_{\langle \alpha m, \beta m' \rangle } \delta_{\sigma\sigma'} ,
\end{eqnarray}
where $\langle \alpha m , \beta m' \rangle$ ($\alpha \not= \beta$) means that $(\alpha, m)$ and $(\beta , m')$ are two intercluster nearest-neightbor sites.
The local density of states (LDOS) at site $i=(\alpha, m)$ is given by the imaginary part of the CPT Green's function,
\begin{equation}
A_{i}(\omega)=-\frac{1}{\pi} \lim_{\eta\rightarrow 0^{+}} \sum_{\sigma}\textrm{Im}\ G_{\mathrm{cpt}, i\sigma, i\sigma}(\omega-\mu+i\eta) ,
\end{equation}
where $\mu$ is the chemical potential and $\eta$ is a broadening parameter.

In our calculation, $\eta=0.24$ is taken. The whole system contains $6\times 6$ clusters. The  periodic boundary conditions are assumed. The parent compound is at half-filling. The single electron is doped into one of the 36 clusters, which contains one more electron than other clusters and is denoted as the cluster $\gamma$.
The results presented below are obtained with $U=10$ if not explicitly specified.

Figure ~\ref{fig1}(a) shows a 12-site cluster tiling that is used in our calculation.
By considering the symmetry of the cluster, we find that the sites with the same color shown in Fig.~\ref{fig1}(a) are equivalent to each other and have identical spectra.
Fig.~\ref{fig1}(b) shows the LDOS on the red and blue sites at half-filling.
The LDOS does not distribute uniformly among the lattice sites within a cluster. This is an artifact of the approximation used in the CPT since the intra- and inter-cluster hoppings are treated differently in this theory.
But the gap values between the lower and upper Hubbard bands and the peak positions, indicated by the grey arrows in Fig.~\ref{fig1}(d-e), are the same on all the sites within a cluster. This is because the on-site Hubbard interaction is treated rigorously in the CPT.

Now let us consider the system with one electron doping. Fig.~\ref{fig1}(c) shows the LDOS for the one-electron doped Hubbard model on the red and blue sites within the doped cluster (for comparison with the result shown in Fig.~\ref{fig1}(b), the zero point of energy is rescaled to the middle of the Mott gap). A broad in-gap state emerges below the upper Hubbard band on the central lattice sites within this doped cluster.
The weight of this broad in-gap state drops quickly on the site moving away from the center of the cluster, indicating that this in-gap state is highly localized around the doping center. The emergence of the in-gap state is accompanied by a reduction in the spectral weights of both the upper and lower Hubbard bands. Fig.~\ref{fig1}(d) shows the difference in the LDOS on the same lattice sites between the $\gamma$ cluster and a cluster $\alpha$ which is far away from the $\gamma$ cluster,
\begin{equation}
\Delta (\omega) = A_{\gamma m} (\omega) - A_{\alpha m}(\omega).
\end{equation}
It shows clearly that the in-gap state results mainly from the spectral transfer from the lower Hubbard band.

\begin{figure}[b]
\begin{center}
\includegraphics[width=0.44\textwidth]{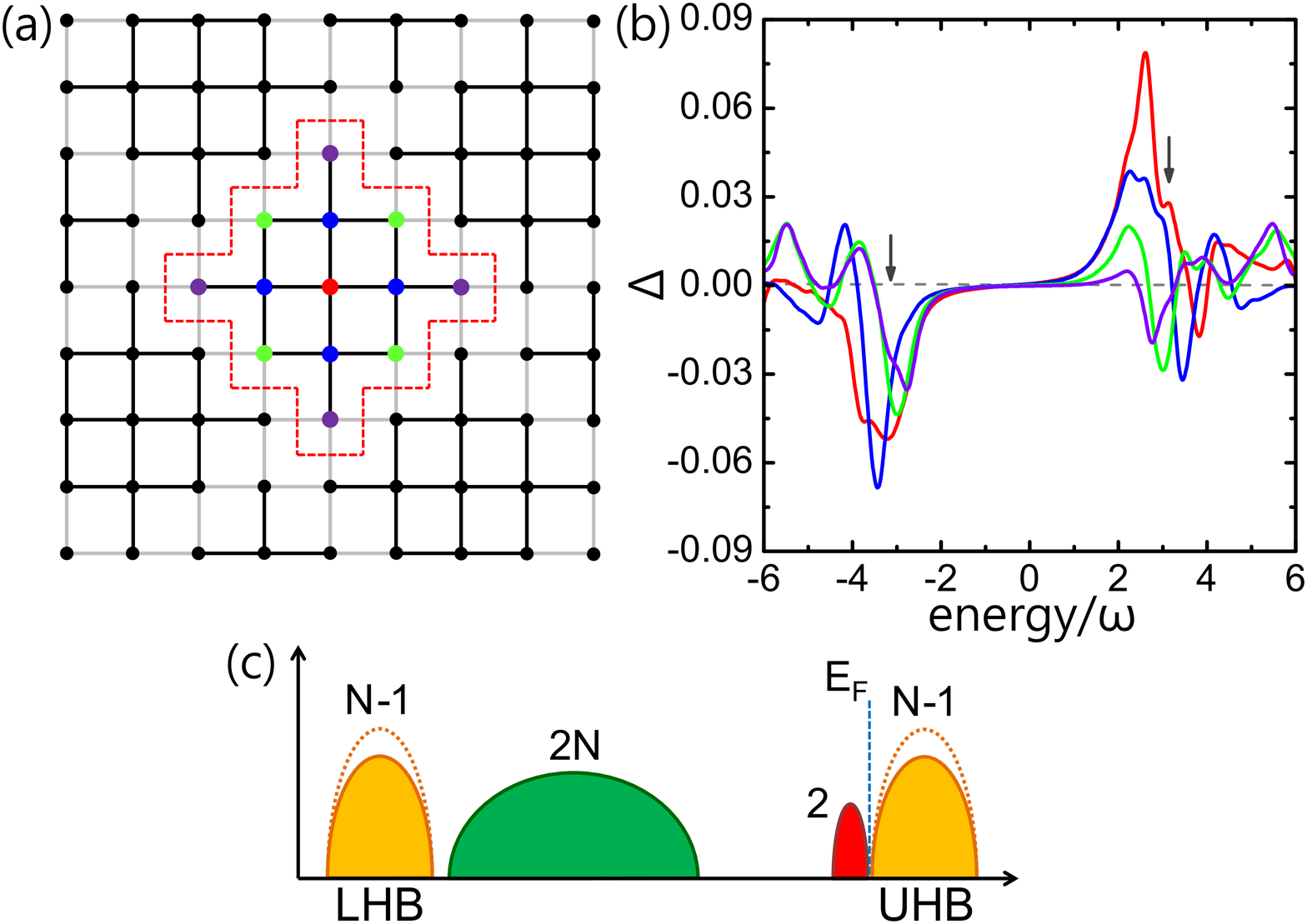}
\caption{(color online)
(a) A 13-site cluster, as enclosed by the red dashed lines, on the square lattice.
(b)$\Delta$ on the sites indicated by the corresponding color sites in (a). The grey arrows indicate the peak positions in the LDOS for the lower and upper Hubbard bands without doping.
(c) Schematic diagram illustrating the spectral weight transfer in the localized limit in a single electron doped three-band model. The green part between the lower (LHB) and upper (UHB) Hubbard bands represents the spectrum from the oxygen charge transfer band.
}\label{fig2}
\end{center}
\end{figure}

Because of the lattice symmetry, the four red sites have the same LDOS. To simulate the recent STM experimental result for $\mathrm{Ca_2CuO_2Cl_2}$ with one Cl vacancy\cite{YY.Wang}, we introduce a small attractive potential $v=-0.08$ to one of the four central sites in the $\gamma$ cluster (i.e. the site in red in the inset of Fig.~\ref{fig1}(e)). As expected, the four central sites in this cluster are no longer equivalent to each other. The in-gap state is well localized around the site with an attractive potential.
This localized feature of the in-gap state has also been found when a 13-site cluster tiling is used on the square lattice (see Fig.~\ref{fig2}(a-b)). It is consistent with the experimental observation.\cite{YY.Wang}

The above result can be understood from the Hubbard model in the localized limit $U /t \rightarrow \infty$. In a $N$-site system at half filling, the total spectral weights for both the lower and upper Hubbard bands are equal to $N$.
Upon one-electron doping, the doped site is doubly occupied.
This eliminates simultaneously one singly occupied state and one channel for adding an electron to a site which is already occupied. Thus the total spectral weights for both the lower and upper Hubbard bands are reduced by 1 and the missing spectral weight is transferred to the bottom of the upper Hubbard band just below the Fermi level~\cite{GA.Sawatzky, H.Eskes, LF.Feiner}, consistent with our calculation.
In the STM measurement~\cite{YY.Wang} on $\mathrm{Ca_2CuO_2Cl_2}$, the spectral weight of the in-gap state seems to come mainly from the upper Hubbard band.
To reconcile this discrepancy between our calculation and the experimental observation, we point out that the energy gap measured by the STM~\cite{YY.Wang} is actually the gap between the upper Hubbard band and the charge transfer band of oxygen whose energy lies between the lower and upper bands, as illustrated by the green spectral weight in Fig.~\ref{fig2}(c).
Thus the spectral weight transfer in the lower Hubbard band, which lies well below the oxygen charge transfer band, is difficult to be observed by the STM.

\begin{figure}
\begin{center}
\includegraphics[width=0.44\textwidth]{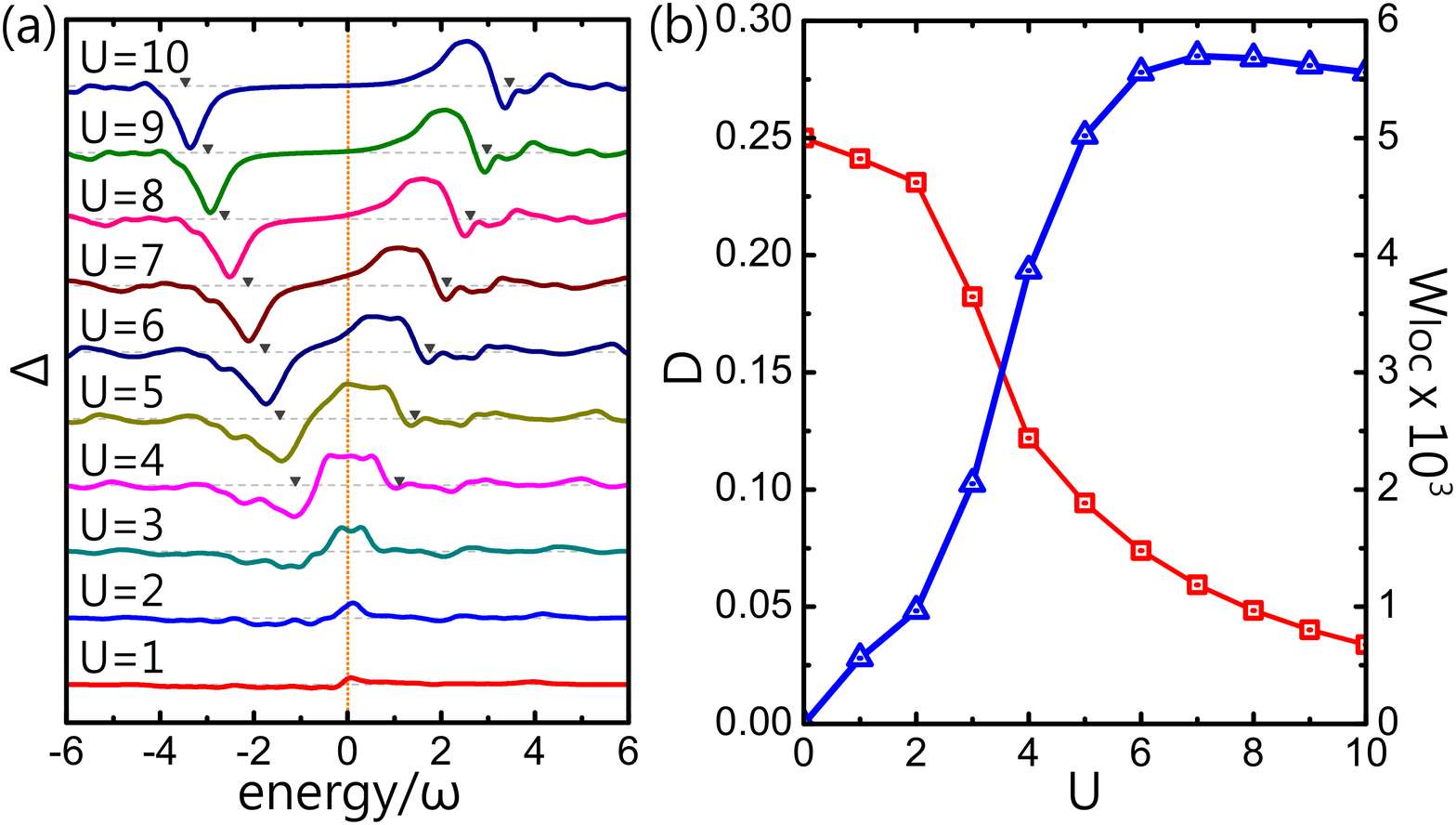}
\caption{(color online)
(a) Difference $\Delta$ in the LDOS  on one of the red sites shown in Fig.~\ref{fig1}(a) for different $U$.
The peak positions of the LDOS for the lower and upper Hubbard bands at half-filling are indicated by the grey triangles.
(b) $U$ dependence of double occupation number $D$ (red line with squares) and the local spectral weight of the in-gap state $W_{\mathrm{loc}}$ (blue line with triangles).
}\label{fig4}
\end{center}
\end{figure}

In order to see how the local in-gap state evolves with the Hubbard interaction, we evaluate the local spectra for several different $U$.
Figure~\ref{fig4}(a) shows the evolution of the difference $\Delta$ in the LDOS for the in-gap state on one of the red colored sites in Fig.~\ref{fig1}(a) with $U$.
From the result, we calculate the total spectral weight for the in-gap state defined by
\begin{equation}
W_{\mathrm{loc}}=\int^{\omega_{2}}_{\omega_{1}}\Delta_{\mathrm{red}}(\omega)\ d\omega
\end{equation}
where $\Delta_{\mathrm{red}}(\omega)$ is the value of $\Delta$ on one of the red colored sites, $\omega_{1}$ and $\omega_{2}$ are the low and high energy boundary of LDOS within which $\Delta_{\mathrm{red}}(\omega)$ is positive and  $\Delta_{\mathrm{red}}(\omega_{1}) = \Delta_{\mathrm{red}}(\omega_{2}) = 0$.
As shown in Fig.~\ref{fig4}(b), $W_{\mathrm{loc}}$ increases rapidly with $U$, especially in the regime $2 < U < 6$, and then becomes nearly saturated above $U \sim 6$.

At half filling, the Mott transition is characterized by a rapid decrease of the double occupation. In order to quantify the correlation between the formation of the local in-gap state and the Mott physics, we calculate the double occupation number $D$ defined by
\begin{equation}
D = \frac{1}{N_{s}} \sum_{i}\langle n_{i\uparrow}n_{i\downarrow}\rangle 
\end{equation}
at half filling.
Fig.~\ref{fig4}(b) shows the double occupation number as a function of $U$.
As expected, $D$ is equal to $0.25$ in the non-interacting limit $U \rightarrow 0$ and drops with increasing $U$. Although it approaches to zero only in the limit $U\rightarrow \infty$, a rapid decrease occurs in the regime of $2 < U < 6$, it coincides with the regime in which $W_{\mathrm{loc}}$ increases rapidly.
This suggests that the local in-gap state is an inherent feature of Mott insulators.

\begin{figure}
\begin{center}
\includegraphics[width=0.44\textwidth]{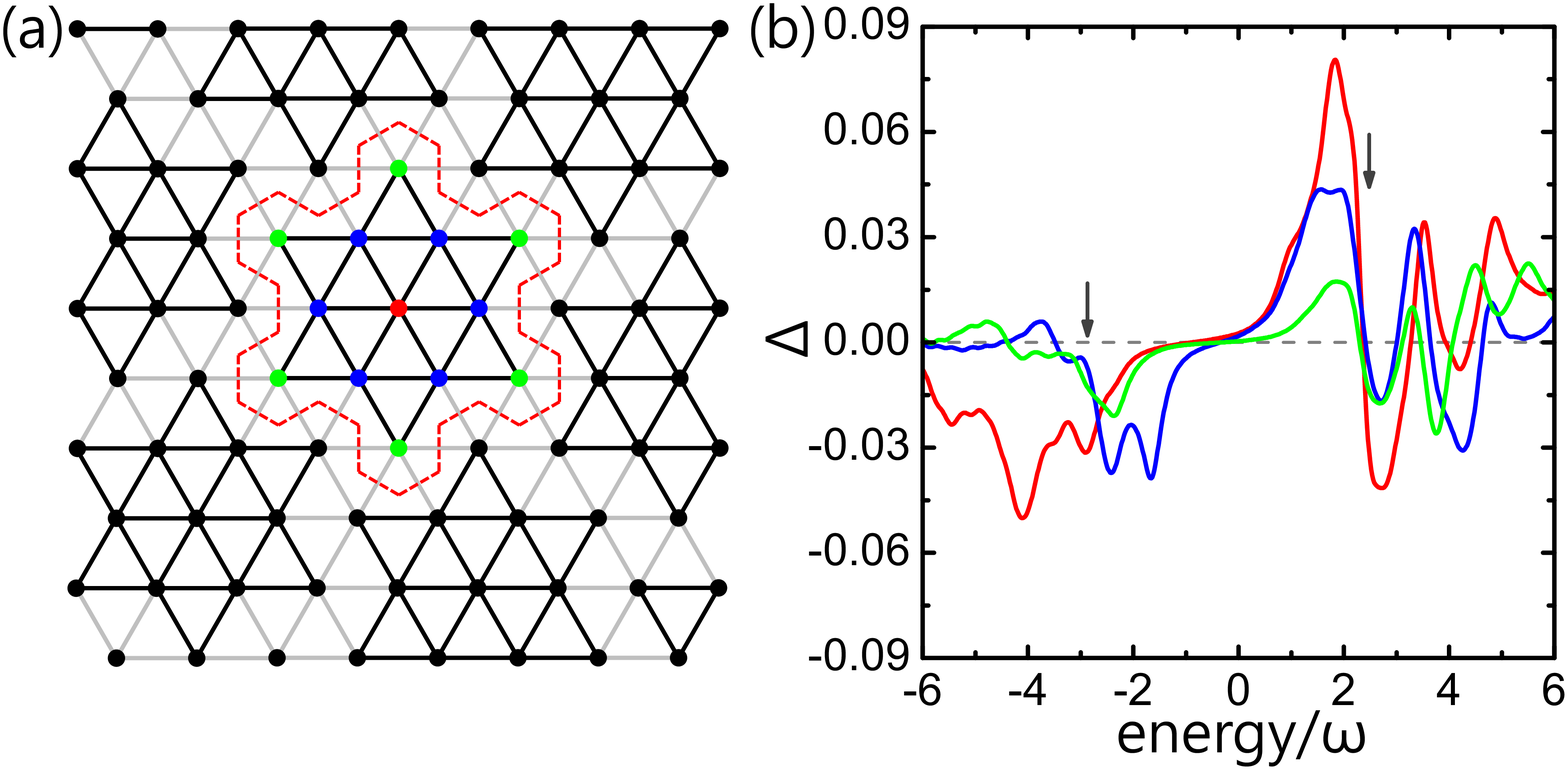}
\caption{(color online)
(a) A 13-site cluster, as enclosed by the red dash lines, on the triangular lattice.
(b) The difference $\Delta$ in the LDOS on the sites with the same colors shown in panel (a).
The peak positions in the LDOS for the lower and upper Hubbard bands at half filling are indicated by the grey arrows.
}\label{fig5}
\end{center}
\end{figure}

To explore the effect of lattice frustration on the localized in-gap state, we have also studied the single electron doped Hubbard model on triangular lattices.
A 13-site hexagram cluster tiling in a system containing 36 clusters with periodic boundary conditions is considered (Fig.~\ref{fig5}(a)) in our calculation.
Figure~\ref{fig5}(b) shows the difference in the LDOS on different sites.
An in-gap state below the upper Hubbard band (indicated by the right grey arrow) appears in the doped cluster and fades away from the doping center.
This is similar to the result found on the square lattice and indicates that the local in-gap state is indeed a result of Coulomb repulsion in a Mott insulator.

\begin{figure}
\begin{center}
\includegraphics[width=0.48\textwidth]{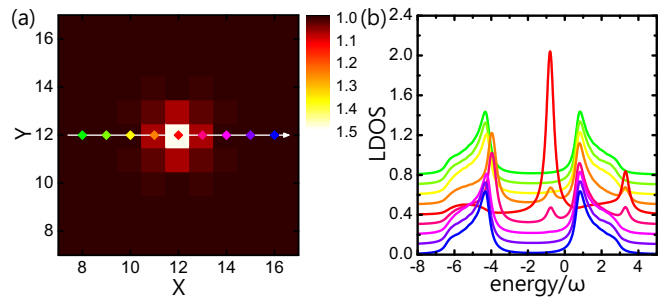}
\caption{(color online).
(a) Real-space distribution of electron occupation numbers obtained by the SCMBA with  $U=6$.
(b) LDOS as a function of energy on the lattice sites shown in the same colors in (a).
An on-site attractive potential $v=-0.08$ on the site (12, 12) has also been added as used in Fig.1(e).
}\label{fig6}
\end{center}
\end{figure}

As mentioned previously, a localized state was found in the one-hole doped Hubbard model by the SCMFA~\cite{WP.Su, AR.Bishop, G.Seibold}. By considering the particle-hole symmetry, one may expect that a localized in-gap state should also exist in the one-electron doped Hubbard model. Now let us make a comparison between the results obtained by the CPT and SCMFA. In the SCMFA, the Hamiltonian is approximated by
\begin{equation}\label{eqmf}
H_{mf}= -t \sum_{\langle ij \rangle} c^\dagger_{i\sigma} c_{j\sigma} +  U\sum_{i\sigma}\langle\hat{n}_{i-\sigma}\rangle \left( 2\hat{n}_{i\sigma}-  \langle\hat{n}_{i\sigma}\rangle \right).
\end{equation}
This mean-field Hamiltonian is solved self-consistently on a $N_{s}=24\times24$ square lattice with $N_s +1$ electrons and periodic boundary conditions, a small attractive potential $v=-0.08$ has also been added to the site (12,12). Figure~\ref{fig6}(a) shows the spatial distribution of the occupation number with $U=6$. We find that the doped electron is highly localized around the site with an attractive potential. The occupation number shows a sharp peak on that site and drops quickly away from it. Figure~\ref{fig6}(b) shows the LDOS on several lattices sites around (12, 12) along the $x$-axis direction. An in-gap state below the upper Hubbard band is clearly seen, in agreement with the CPT calculation, although the spectral weight of the in-gap state at the central site is overestimated by the SCMFA. It suggests that the localized in-gap state is an intrinsic property of the one-electron Hubbard model, and it exists independent of the approximations used in our calculation.

In summary, we calculate the local density of states for the Hubbard model at half filling or with one electron doping on both square and triangular lattices using CPT.
It is found that an in-gap state below the upper Hubbard band exists at and near the doping center irrespective of the lattice structure. The spectral weight of this in-gap state is inherently anti-correlated with the double occupation number of electrons, which is a key variable characterizing a Mott insulator, in the relevant physical parameter range. It indicates that the in-gap state is driven by the strong correlation effect associated with the Mott transition. Our result gives a natural account for the STM experimental observation~\cite{YY.Wang}.

\begin{acknowledgments}
This work was supported by the National Natural Science Foundation of China (Grants No.11190024, No.10934008, No.11374138 and No.11204125) and by the National Basic Research Program of China (Grant No.2011CB309703,  No. 2011CB922101 and No.2011CB605902).
\end{acknowledgments}

\end{document}